\documentclass[aps,prd,twocolumn,superscriptaddress,10pt,nofootinbib,floatfix]{revtex4-2}
\usepackage{eurosym}
\usepackage{amssymb}
\usepackage{textcomp}
\usepackage{graphicx}
\usepackage{epsfig}
\usepackage{subfigure}
\usepackage{amsmath}
\usepackage{slashed}
\usepackage{units}
\usepackage{array}
\usepackage{lmodern}
\usepackage{dcolumn}
\usepackage{bm}
\usepackage{cancel}

\begin{document}

\title{Braneworld Baryogenesis and QCD-Era Magnetogenesis: A Predictive Link}
\author{Micha\"{e}l Sarrazin}
\email{michael.sarrazin@ac-besancon.fr}
\affiliation{Universit\'{e}
Marie et Louis Pasteur, CNRS, Institut UTINAM (UMR 6213), \'{E}quipe de Physique Th\'{e}orique, F-25000 Besan\c con, France}

\begin{abstract}
We demonstrate that primordial magnetic fields (PMF) play a decisive role in
the braneworld baryogenesis scenario of [Phys. Rev. D \textbf{110}, 023520
(2024)], where C/CP violation arises from the coupling of visible and hidden
matter-antimatter sectors through a pseudo-scalar field. Although this
mechanism generates baryon number efficiently only after the quark-hadron
transition, by incorporating a realistic stochastic PMF within a
semi-analytical framework, we find that matching the observed
baryon-antibaryon asymmetry robustly requires PMF strengths of order $%
10^{10} $~T right after the transition, in agreement with causal QCD-era
magnetogenesis. We further reveal that magnetic fluctuations drive the
baryon-density spectrum to white noise on large scales, yielding an
isocurvature component compatible with Cosmic Microwave Background (CMB)
bounds. This establishes a predictive link between the braneworld
baryogenesis model and realistic early-Universe magnetic fields.
\end{abstract}

\maketitle

\section{Introduction}

\label{intro}

The origin of the observed matter--antimatter asymmetry remains one of the
central open questions in cosmology and particle physics \cite%
{Bodeker2021,Canetti2012,Cline2019}. Any successful baryogenesis mechanism
must satisfy the Sakharov conditions \cite{Sakharov1967} while remaining
compatible with the thermal history of the early Universe. Among the various
ingredients potentially relevant for baryogenesis, primordial magnetic
fields (PMF) occupy a particularly interesting position. They are
generically expected across a wide range of scenarios -- from inflationary
magnetogenesis \cite{Turner1988,Ratra1992} to causal production during the
electroweak or QCD epochs \cite{Cheng1994,Grasso2001,Subramanian2016} -- and
may influence several key cosmological processes, from structure formation 
\cite{Widrow2002,Widrow2012} to baryon-number generation \cite%
{Joyce1997,Grasso1998,Long2014,Fujita2016,Ralegankar2025}.

In a recent work \cite{Sarrazin2024}, we proposed a novel baryogenesis
mechanism arising in a two-brane Universe described at low energies by a
noncommutative spacetime $M_{4}\times Z_{2}$ \cite%
{Sarrazin2010,Stasser2019,Stasser2020}. In this framework, C/CP violation
emerges naturally through a pseudo-scalar field derived from the interbrane $%
U(1)_{+}\otimes U(1)_{-}$ electromagnetic gauge structure. The mechanism
becomes efficient once primordial magnetic fields generate a phase
difference between the gauge potentials on the two branes, thereby inducing
asymmetric neutron--hidden-neutron transitions. It was shown in Ref. \cite%
{Sarrazin2024} that this process, triggered immediately after the
quark--gluon plasma to hadron gas (QGP--HG) transition, can reproduce the
observed baryon-antibaryon asymmetry.

However, the PMF considered in Ref.~\cite{Sarrazin2024} was modeled as a
single fixed configuration. While motivated by the QCD era, this approach
remained \textit{ad hoc} and did not incorporate realistic magnetogenesis
scenarios. As a consequence, two key questions remained open: (i) how the
baryogenesis mechanism behaves when embedded in a realistic stochastic PMF;
and (ii) whether the PMF amplitude required to reproduce the observed
asymmetry is compatible with those expected from causal QCD-scale
magnetogenesis.

The purpose of the present work is to address these questions by
incorporating a statistically realistic description of primordial magnetic
fields into the two-brane baryogenesis framework.

Instead of prescribing a fixed configuration, we model the PMF as a causal
stochastic field with a broken power-law spectrum motivated by QCD-scale
plasma turbulence and instabilities (bubble or domain-wall dynamics being
relevant only in scenarios that render the QCD transition first order) \cite%
{Brandenburg1996,Forbes2000,Boyanovsky2005,Boeckel2012,Tevzadze2012}. This
allows us to propagate the full distribution of magnetic fluctuations
through the non-linear interbrane transition dynamics derived from Ref. \cite%
{Sarrazin2024}.

This leads to two central results.

First, the amplitude of the primordial magnetic field is no longer an input
parameter but becomes a required prediction of the baryogenesis dynamics. We
show that reproducing the observed baryon density robustly requires PMF
strengths of order $B\sim 10^{10}\ $T, at the QCD epoch \cite%
{Brandenburg1996,Forbes2000,Boyanovsky2005,Boeckel2012,Tevzadze2012}.
Remarkably, this value coincides with predictions from causal QCD
magnetogenesis mechanisms involving plasma turbulence and instabilities
during the QGP--HG transition, as well as bubble dynamics in scenarios
featuring a first-order transition \cite%
{Brandenburg1996,Forbes2000,Boyanovsky2005,Boeckel2012,Tevzadze2012}. This
agreement is highly non-trivial: it indicates that the exotic brane-induced
C/CP-violating dynamics required for baryogenesis naturally select the same
magnetic-field amplitude produced by standard QCD plasma physics. The PMF
amplitude is thus not tuned but emerges as a predictive link between
beyond-Standard-Model baryogenesis and standard microphysics.

Second, spatial fluctuations of the magnetic vector potential induce
baryon-density inhomogeneities. Using a semi-analytical two-point method, we
show that the resulting baryon-density power spectrum is universally white
noise on large scales ($P_{\delta }(k)\propto k^{0}$), independently of the
PMF spectral index. This universality originates from the short-range
correlations of a causal PMF combined with the strongly non-linear mapping
between the magnetic potential and baryon-number production. The
corresponding baryon isocurvature mode is statistically independent of the
primordial adiabatic fluctuations and its amplitude lies far below current
CMB bounds \cite{Planck2018}. While not an observable signature, it
constitutes a robust internal prediction of the scenario and establishes its
compatibility with existing cosmological constraints.

It is instructive to contrast our mechanism with existing PMF-assisted
baryogenesis scenarios \cite%
{Joyce1997,Grasso1998,Long2014,Fujita2016,Ralegankar2025}. For instance, in
the leading models \cite{Fujita2016} the baryon asymmetry is produced from
helical magnetic fields through the chiral anomaly of the Standard-Model $%
U(1)_{Y}$ interaction: a time-varying magnetic helicity sources baryon
number directly, so that the field acts as a reservoir of helical charge. In
such scenarios helicity is indispensable -- a non-helical field sources no
global baryon number and yields no asymmetry. Our mechanism differs
qualitatively. The magnetic field does not supply baryon number through an
anomaly; rather, its vector potential sets the phase $\theta $ that drives
the C/CP-violating interbrane neutron--hidden-neutron transitions \cite%
{Sarrazin2024}. The field therefore acts as a trigger rather than as a
charge reservoir, and helicity plays no essential role -- indeed our
primary, non-helical cases ($n=0$ and $n=4$) already reproduce the observed
asymmetry. This makes the present scenario a genuinely distinct route by
which primordial magnetism can seed the baryon asymmetry, complementary to
the anomaly-based mechanisms.

The rest of the paper is organized as follows. Section~\ref{II} outlines the
two-brane baryogenesis mechanism developed in Ref. \cite{Sarrazin2024}, but
we also compute for the first time the baryon density against the strength
of the magnetic vector potentials in the primordial universe. Section~\ref{III}
presents the stochastic model of primordial magnetic fields under
consideration. Section~\ref{methodology} describes the semi-analytical method of
computation of the baryon-density spectrum. Finally, Section~\ref{sec:results} discusses the
magnetic-field amplitude required by baryogenesis in light of our results,
as well as the universal emergence of a white-noise spectrum and its
compatibility with CMB isocurvature limits.

\section{Two-Brane Baryogenesis Model}

\label{II}

\subsection{Theoretical Framework of the Present Study}

\label{IIA}

For completeness and to make the present paper self-contained, we summarize
below the essential features of the two-brane baryogenesis mechanism
introduced in Ref. \cite{Sarrazin2024}. This mechanism relies on the
equivalence between a two-brane universe embedded in a $(3+N,1)$-dimensional
bulk and a non-commutative two-sheeted space-time $M_{4}\times Z_{2}$, when
dealing with quantum dynamics of fermions and their related gauge fields 
\cite{Sarrazin2010,Stasser2019,Stasser2020}. This equivalence is not a
phenomenological \textit{ansatz}, it has been demonstrated in previous works 
\cite{Sarrazin2010,Stasser2019,Stasser2020} and applies broadly to
braneworld theories, from string-inspired models to domain walls frameworks.

In a two-brane universe, our visible Universe is a 3-brane coexisting with a
hidden 3-brane in a $(3+N,1)$-dimensional bulk ($N\geq 1$). At low energies,
this system is described by a non-commutative two-sheeted space-time $%
M_{4}\times Z_{2}$, with fermion dynamics governed by the Lagrangian 
\begin{equation}
\mathcal{L}_{M_{4}\times Z_{2}}\sim \bar{\Psi}(i\cancel{D}-m)\Psi ,
\label{Lag}
\end{equation}%
where $\Psi =(\psi _{+},\psi _{-})^{T}$ contains the wave functions on the
visible (+) and hidden (-) branes, $m$ is the fermion mass, and $\cancel{D}%
=\Gamma ^{N}D_{N}=\Gamma ^{\mu }D_{\mu }+\Gamma ^{5}D_{5}$ is the Dirac
operator acting on $M_{4}\times Z_{2}$. The derivative operators acting on $%
M_{4}$ and $Z_{2}$ are $D_{\mu }=\mathbf{1}_{8\times 8}\partial _{\mu }$ ($%
\mu =0,1,2,3$) and$\ D_{5}=ig\sigma _{2}\otimes \mathbf{1}_{4\times 4}$,
respectively, including a bare coupling constant $g$ between branes. The
gamma matrices are defined as: $\Gamma ^{\mu }=\mathbf{1}_{2\times 2}\otimes
\gamma ^{\mu }$\ and\ $\Gamma ^{5}=\sigma _{3}\otimes \gamma ^{5}$, where $%
\gamma ^{\mu }$ and $\gamma ^{5}=i\gamma ^{0}\gamma ^{1}\gamma ^{2}\gamma
^{3}$ are the usual Dirac matrices and $\sigma _{k}$ ($k=1,2,3$) the Pauli
matrices. Equation (\ref{Lag}) is characteristic of fermions in
non-commutative $M_{4}\times Z_{2}$ two-sheeted space-times as introduced by
other authors \cite{Connes1990,Lizzi1997,Kase1999,Kase2001}.

A pseudo-scalar Higgs-like field $\phi $, emerging from the $U(1)_{+}\otimes
U(1)_{-}$ electromagnetic gauge field in the two-brane universe \cite%
{Sarrazin2010,Stasser2019,Stasser2020}, couples to fermions, yielding a
global coupling \cite{Sarrazin2024}: 
\begin{equation}
g\rightarrow g+iq\phi _{0},\quad \phi _{0}=\eta (e^{i\theta }-i),
\label{coupl}
\end{equation}%
where $\phi _{0}$ denotes the vacuum state, with $\eta =g/e$, $e$ is the
elementary charge, $q$ the fermion charge (quark or lepton), and $\theta $
the scalar field phase, driven by the magnetic vector potential difference 
\cite{Sarrazin2024} 
\begin{equation}
\theta =e\int \left( A_{\mu }^{+}-A_{\mu }^{-}\right) dx^{\mu },
\label{eq:theta}
\end{equation}%
where $A_{\mu }^{\pm }$ are the electromagnetic potentials in each brane. We
stress that the phase $\theta $ defined by Eq.~(\ref{eq:theta}) is the phase
of the \textquotedblright order parameter\textquotedblright\ associated with
the relative $U(1)_{+}\otimes U(1)_{-}$ gauge structure of the two-brane
system, i.e. a Goldstone mode \cite{Sarrazin2024}. As such, Eq.~(\ref%
{eq:theta}) is gauge-covariant rather than gauge-invariant: under $A_{\mu
}^{\pm }\rightarrow A_{\mu }^{\pm }+\partial _{\mu }\Lambda _{\pm }$ one has 
$\theta \rightarrow \theta +e(\Lambda _{+}-\Lambda _{-})$, exactly as for an
open Wilson line or an Aharonov--Bohm phase along an open path (the full
derivation is given in Ref.~\cite{Sarrazin2024}). The phase $\theta $ is
therefore not by itself a physical observable, and it need not be. Now, it
can be shown \cite{Sarrazin2024} that the effective interbrane coupling
constant for fermions of the Standard Model becomes 
\begin{equation}
g\rightarrow \mathfrak{g}=g\sqrt{1+2z(1+z)(1-\sin \theta )}  \label{ggot}
\end{equation}%
where $z=q/e$, while for anti-fermions, $\bar{\mathfrak{g}}$ simply differs
due to charge conjugation (i.e. $\bar{\mathfrak{g}}=\mathfrak{g}(-z)$) \cite%
{Sarrazin2024}.

From Eqs. (\ref{Lag}) to (\ref{eq:theta}), and using a quark constituent
description \cite{Sarrazin2010,Stasser2019,Stasser2020} of baryons, it was
demonstrated that the relevant neutron-hidden neutron transitions ($%
n\leftrightarrow n^{\prime }$) are described by the Hamiltonian \cite%
{Sarrazin2024} 
\begin{equation}
\mathcal{W}=\varepsilon 
\begin{pmatrix}
0 & \mathbf{u} \\ 
\mathbf{u}^{\dagger } & 0%
\end{pmatrix}%
,\quad \varepsilon =\mathfrak{g}\mu |\mathbf{A}_{+}-\mathbf{A}_{-}|,
\label{W}
\end{equation}%
with $\mu $ the neutron magnetic moment, $\mathbf{u}$ a unitary matrix and
where $\mathfrak{g}$ (see Eqs. (\ref{gmean}) and (\ref{gbmean}) in the
following subsection) was here building for the neutron from the quark
constituent model \cite{Sarrazin2010,Stasser2019,Stasser2020}. $\mathbf{A}%
_{\pm }$ are the magnetic vector potentials related to the ambient magnetic
fields $\mathbf{B}_{\pm }=\nabla \times \mathbf{A}_{\pm }$ in each brane.
Roughly, the transition rate between branes is $\gamma \sim 2\varepsilon
^{2}/\Gamma $, where $\Gamma $ is the collisional rate between neutrons and
other particles species during the Big Bang \cite{Sarrazin2024} \footnote{%
Note that the gauge invariance of the mechanism is carried by the transition
energy $\varepsilon $ of the Hamiltonian (\ref{W}): although $\varepsilon $
is written compactly as $\varepsilon =\mathfrak{g}\,\mu \,|\mathbf{A}_{+}-%
\mathbf{A}_{-}|$ in terms of the gauge-covariant quantities $\mathfrak{g}%
(\theta )$ and $|\mathbf{A}_{+}-\mathbf{A}_{-}|$, the physical transition
amplitude built from it -- and hence the rate $\gamma \sim 2\varepsilon
^{2}/\Gamma $ and the resulting asymmetry $Y_{B}-Y_{\overline{B}}$ -- is
exactly gauge invariant. This invariance is not manifest at the level of
Eq.~(\ref{W}); exhibiting it explicitly requires the full derivation of the
interbrane transition within the quark-constituent description, given in
Ref.~\cite{Sarrazin2024}. We do not reproduce that calculation here and
simply rely on its result: the predictions of the model are gauge invariant
by construction. As discussed in Sec.~\ref{IIB}, the unspecified
integration path in Eq.~(\ref{eq:theta}) is then handled through a
statistical average over $\theta $, which yields the effective 
couplings directly.}. For antineutrons, $\bar{\mathfrak{g}}$ (as $\bar{\mathfrak{g}}$ $%
\neq \mathfrak{g}$) modifies transitions to the hidden brane, inducing C/CP
violation \cite{Sarrazin2024}. The resulting baryon-antibaryon asymmetry is
computed using generalized Lee-Weinberg equations, derived from Boltzmann
transport equations (see Ref. \cite{Sarrazin2024} for details; the related 
\texttt{COMPUBARYO2B.py} code is provided in the \textquotedblright Data and
Code Availability\textquotedblright\ section of the present paper), which
describe the evolution of comoving density $Y_{B}=n_{B}/s$ and $Y_{\overline{%
B}}=n_{\overline{B}}/s$ (for baryons and antibaryons respectively) relative
to the entropy density $s$ in both branes. These equations account for
baryon-antibaryon annihilation and interbrane transitions, yielding $%
Y_{B}-Y_{\overline{B}}$ to be consistent \cite{Sarrazin2024} with
observations ($Y_{B}-Y_{\overline{B}}=(8.8\pm 0.6)\times 10^{-11}$) \cite%
{Cline2019}. This mechanism \cite{Sarrazin2024}, active right after the
QGP-HG transition ($T\approx 160$ MeV) thanks to the PMF \cite%
{Brandenburg1996,Forbes2000,Boyanovsky2005,Boeckel2012,Tevzadze2012},
satisfies the Sakharov conditions \cite{Sakharov1967}, meaning both branes
must undergo two different temperatures. Then, setting $T$ as the
temperature in our visible braneworld and $T^{\prime }$ in the hidden
braneworld, one defines 
\begin{equation}
\kappa =\frac{T}{T^{\prime }},  \label{kappa}
\end{equation}%
where $\kappa $ is a constant parameter \cite{Sarrazin2024}. At the level of
the background cosmology, the two branes are assumed to share a common
Friedmann--Lema\^{\i}tre--Robertson--Walker (FLRW) geometry, allowing for
small departures in their respective thermal histories. This assumption
requires that any relative difference in energy density remains
perturbative, 
\begin{equation}
\frac{\delta \rho }{\rho }\ll 1.  \label{cond1}
\end{equation}%
In practice, cosmological perturbation theory remains under control as long
as $\delta \rho /\rho \lesssim 10^{-1}$, beyond which quadratic
back-reaction effects become non-negligible and the notion of a single
effective background expansion ceases to be well defined \cite%
{Mukhanov2005,Weinberg2008}. Since the energy density of a relativistic
plasma scales as $\rho \propto T^{4}$, this implies a conservative upper
bound on the relative temperature mismatch, 
\begin{equation}
\frac{\left\vert \Delta T\right\vert }{T}\lesssim 10^{-2},  \label{cond2}
\end{equation}%
i.e. 
\begin{equation}
\left\vert \kappa -1\right\vert \lesssim 10^{-2}.  \label{cond3}
\end{equation}%
Throughout this work, we therefore restrict attention to this perturbative
regime, in which the cosmological background remains well described by a
common FLRW evolution. We note that the small relative temperature mismatch
assumed in this work should be regarded as an effective initial condition
rather than the outcome of a specific microscopic mechanism. In a
brane--collision scenario, a strictly vanishing temperature difference would
require an exact symmetry between the two branes, including identical
tensions, couplings to bulk degrees of freedom and perfectly synchronized
collision dynamics. Such an exact symmetry is non-generic in non-equilibrium
cosmological settings. By contrast, small temperature asymmetries with $%
\left\vert \Delta T\right\vert /T\ll 1$ naturally arise in brane--collision
scenarios from mild geometric asymmetries, slightly different couplings to
bulk fields, or non-simultaneous energy transfer during the collision \cite%
{Khoury2001,Steinhardt2001}. A detailed dynamical origin of this asymmetry
lies beyond the scope of the present work and is left for future
investigation.

\subsection{Expression of the Scalar Field Phase and Effective Interbrane
Coupling Constants}

\label{IIB}

The scalar field phase $\theta $, central to our two-brane baryogenesis
model, is defined by Eq. (\ref{eq:theta}) (see also Eq. (14) in our previous
work \cite{Sarrazin2024}). At the QGP-HG transition, plasma dynamics
suppress the temporal components $A_{0}^{\pm }\approx 0$ \cite%
{Baym1996,Banerjee2004}, leaving the spatial components $\mathbf{A}_{\pm }$
dominant. The unspecified integration path in Eq. (\ref{eq:theta})
necessitates a statistical treatment to be treated with the Boltzmann
transport equations (equations 47--50 of \cite{Sarrazin2024}) used to
compute the baryon-antibaryon asymmetry $Y_{B}-Y_{\overline{B}}$.

To capture the quantum dynamics of the pseudo-scalar field $\phi =\eta
(e^{i\theta }-i)$, we define $\theta $ as a path-dependent quantity in a
Feynman path integral framework. For a path $\gamma _{\text{path}}$ from an
initial space-time point $x_{0}=(t_{0},\mathbf{x}_{0})$ to a final point $%
x=(t,\mathbf{x})$, the phase is 
\begin{equation}
\theta \lbrack \gamma _{\text{path}}]=e\int_{\gamma }(\mathbf{A}_{+}-\mathbf{%
A}_{-})\cdot d\mathbf{l},  \label{func}
\end{equation}%
where the integral is spatial due to $A_{0}^{\pm }=0$. The quantum average
of $\theta $ is 
\begin{equation}
\langle \theta \rangle =\frac{\int \mathcal{D}[\gamma _{\text{path}%
}]\,\theta \lbrack \gamma _{\text{path}}]\,e^{iS[\gamma _{\text{path}%
}]/\hbar }}{\int \mathcal{D}[\gamma _{\text{path}}]\,e^{iS[\gamma _{\text{%
path}}]/\hbar }},  \label{funcav}
\end{equation}%
where $\mathcal{D}[\gamma _{\text{path}}]$ is the functional measure over
paths, and $S[\gamma _{\text{path}}]$ is the action of the scalar field $%
\varphi =2e^{-i\theta }(\phi -\phi _{0})$ describing the fluctuations of the
scalar field $\phi $ around the vacuum state, and governed by the Lagrangian
(see equation (15) of \cite{Sarrazin2024}) 
\begin{equation}
\mathcal{L}=\frac{1}{2}(\partial _{\mu }\varphi )(\partial ^{\mu }\varphi )-%
\frac{1}{2}m_{\varphi }^{2}\varphi ^{2},\quad m_{\varphi }=2g.
\label{LagScal}
\end{equation}

In the pure vacuum state, fluctuations of $\varphi $ are negligible ($%
\varphi \ll \eta $, see Sec. V of \cite{Sarrazin2024}), and $\theta $
behaves as a Goldstone mode. In an isotropic and homogeneous universe, the
vectorial contributions of $\mathbf{A}_{+}-\mathbf{A}_{-}$ cancel over all
path directions 
\begin{equation}
\langle \theta \lbrack \gamma _{\text{path}}]\rangle \approx e\int \frac{%
d\Omega }{4\pi }(\mathbf{A}_{+}-\mathbf{A}_{-})\cdot \mathbf{n}|\mathbf{x}-%
\mathbf{x}_{0}|=0,  \label{path}
\end{equation}%
where $\mathbf{n}$ is the unit path vector and $d\Omega $ the solid angle.
Quantum interference across paths further enforces $\langle \theta \rangle
=0 $, which we adopt as the primary approximation for numerical
calculations. This vanishing of the mean value is a direct consequence of
the isotropy of the early universe and fixes the centering of the
distribution of $\theta $. It is not, however, sufficient by itself to
determine the full distribution: a vanishing first moment is compatible with
many non-uniform laws. The uniformity of $\theta$ modulo $2\pi $, rather
than that of $\theta$ itself, is the relevant property for the interbrane
couplings derived from (\ref{ggot}) and results from the magnitude of the
accumulated phase.

Along any quantum-coherent segment of length $\ell $, Eq.~(\ref{func}) gives 
$\theta \sim (e/\hbar )\,A_{t}\,\ell $, with a prefactor $eA_{t}/\hbar
\approx 6\times 10^{23}~\mathrm{m^{-1}}$, obtained directly from the typical
amplitude $A_{t}\sim \mathrm{few}\times 10^{8}~\mathrm{T\,m}$ of the
magnetic vector potential required by baryogenesis at the QCD epoch (see
Table~\ref{tab:a0_values} and Sec.~\ref{VA}) \footnote{%
We stress that this amplitude is fixed by the vector potential itself, not
by the naive horizon estimate $A_{0}\sim cB_{0}H^{-1}$ ($H$ is the Hubble
constant) used in our previous work \cite{Sarrazin2024}: as shown in Sec. %
\ref{IIIB} (see Eq.~(\ref{Bo})), the field strength obeys $B_{0}\sim \alpha
\,A_{0}\,k_{\ast }$ and is therefore controlled by the injection scale $%
k_{\ast }\gg k_{\mathrm{IR}}$, so that $B_{0}\gg HA_{0}/c$ and the present
winding estimate is consistent with the comparatively large field $B_{0}\sim
10^{10}~\mathrm{T}$ used throughout.}. A full $2\pi $ winding of $\theta $
therefore requires only a path length $\ell \sim 10^{-23}~\mathrm{m}$, eight
orders of magnitude below a proton radius. Since any physical coherence
scale in the dense, highly collisional post-QCD plasma vastly exceeds this
bound \cite{Banerjee2004}, the phase $\theta $ winds an enormous number of
times across $\ell $, and $\theta \bmod2\pi $ is extremely close to a
uniform distribution over $[0,2\pi ]$. The result is robust, depending not
on the precise coherence length but only on the (overwhelmingly satisfied)
condition that it exceed $10^{-23}~\mathrm{m}$. 

Now, the coupling constants $\mathfrak{g}$ and $\overline{\mathfrak{%
g}}$ for neutrons and antineutrons, given by equations (31) and (32) of \cite%
{Sarrazin2024}, depend on $\theta \lbrack \gamma _{\text{path}}]$ 
\begin{equation}
\frac{\mathfrak{g}(\theta \lbrack \gamma _{\text{path}}])}{g}=\frac{2}{9}%
\sqrt{5+4\sin \theta \lbrack \gamma _{\text{path}}]}+\frac{1}{9}\sqrt{%
29-20\sin \theta \lbrack \gamma _{\text{path}}]},  \label{gmean}
\end{equation}%
and%
\begin{equation}
\frac{\overline{\mathfrak{g}}(\theta \lbrack \gamma _{\text{path}}])}{g}=%
\frac{2}{9}\sqrt{17-8\sin \theta \lbrack \gamma _{\text{path}}]}+\frac{1}{9}%
\sqrt{5+4\sin \theta \lbrack \gamma _{\text{path}}]}.  \label{gbmean}
\end{equation}

Then, having established the uniformity of $\theta $ modulo $2\pi $, we
compute the effective coupling constants by averaging over all paths 
\begin{eqnarray}
\langle \mathfrak{g}\rangle  &=&\frac{\int \mathcal{D}[\gamma _{\text{path}}]%
\mathfrak{g}(\theta \lbrack \gamma _{\text{path}}])\,e^{iS[\gamma _{\text{%
path}}]/\hbar }}{\int \mathcal{D}[\gamma _{\text{path}}]\,e^{iS[\gamma _{%
\text{path}}]/\hbar }}  \label{gmeanbis} \\
&\sim &\frac{1}{2\pi }\int_{0}^{2\pi }\mathfrak{g}(\theta )\,d\theta \approx
1.0507g,  \notag
\end{eqnarray}%
and with the same for $\overline{\mathfrak{g}}$%
\begin{equation}
\left\langle \overline{\mathfrak{g}}\right\rangle \approx 1.1392g.
\label{gmeanbis2}
\end{equation}%
The non-linear dependence on $\sin \theta $ ensures $\langle \mathfrak{g}%
\rangle \neq \langle \overline{\mathfrak{g}}\rangle $, as terms like $\sqrt{%
5+4\sin \theta }$, $\sqrt{29-20\sin \theta }$, and $\sqrt{17-8\sin \theta }$
yield distinct contributions despite $\langle \sin \theta \rangle =0$. This
sustains a non-zero asymmetry driving baryogenesis. Note that, as discussed
in Sec.~\ref{IIA}, the gauge invariance of the physical predictions is
already guaranteed by the structure of the interbrane Hamiltonian (\ref{W}),
independently of the present averaging: the average over $\theta $ is a
computational device, not the origin of the invariance. What the uniformity
of $\theta $ modulo $2\pi $ provides is the practical means to evaluate the
effective couplings: since the realized distribution is uniform, the
computation reduces to an integration of $\mathfrak{g}(\theta )$ and $%
\overline{\mathfrak{g}}(\theta )$ over $[0,2\pi ]$. This is moreover
consistent with the underlying gauge structure, the uniform measure being
the unique law invariant under $\theta \rightarrow \theta +e(\Lambda
_{+}-\Lambda _{-})$; but it is that underlying invariance, established in
Ref.~\cite{Sarrazin2024}, and not the choice of measure, that makes the
predictions physical. The averaged couplings $\langle \mathfrak{g}\rangle
\approx 1.0507g$ and $\langle \overline{\mathfrak{g}}\rangle \approx 1.1392g$
(see Eqs.~(\ref{gmeanbis}) and (\ref{gmeanbis2})) follow accordingly, 
and are used in the following for the numerical computations.

\subsection{Numerical Calculation of the Baryon Density}

\label{IIC}

While the coupling constants $\mathfrak{g}$ and $\overline{\mathfrak{g}}$
are averaged, the local C/CP violation is driven by the local magnitude of
the potential difference $A_{t}=\left\vert \mathbf{A}_{t}\right\vert =|%
\mathbf{A}_{+}-\mathbf{A}_{-}|$ (see Eq. (\ref{W})), whose spatial
fluctuations are seeded by the stochastic PMF. It is these fluctuations that
source the variations in the final baryon density. As explained here above,
in our previous work \cite{Sarrazin2024}, we developed a numerical code to
compute the baryon-antibaryon asymmetry $Y_{B}-Y_{\overline{B}}$, which
depends on $A_{t}=|\mathbf{A}_{+}-\mathbf{A}_{-}|$ in agreement with known
models of primordial magnetic fields \cite{Cheng1994} -- but without
detailed description of their features \cite{Sarrazin2024}. Then, in this
code, the magnetic fields, and thus $A_{t}$, were supposed to be constant in
the whole universe, and fluctuations were not considered.

\begin{figure}[th]
\centerline{\ \includegraphics[width=8.5cm]{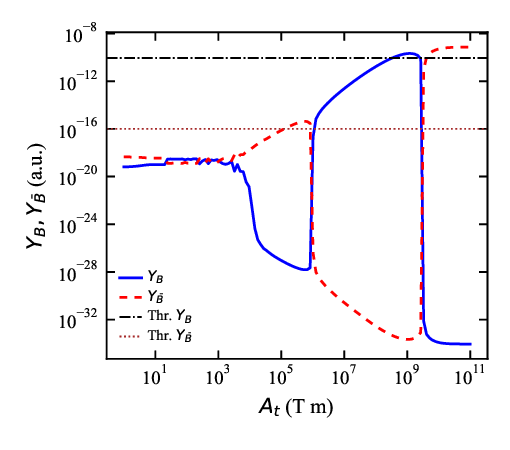}}
\caption{Baryon ($Y_{B}$ : blue solid line) and antibaryon ($Y_{\overline{B}%
} $ : red dashed line) comoving densities at $T=20$ MeV against magnetic
vector potential $A_{t}$ at $T=160$ MeV for $\protect\kappa =0.99$ (i.e. $%
\left\vert \Delta T\right\vert /T\lesssim 10\times 10^{-3})$. Horizontal
black dash-dotted line: observed baryon comoving density ($8.8\times
10^{-11} $). Horizontal red dotted line: upper limit on the expected
antibaryon comoving density ($10^{-16}$).}
\label{f1}
\end{figure}
In the present work, running the code by using values of $\mathfrak{g}$ and $%
\overline{\mathfrak{g}}$ given by Eqs. (\ref{gmeanbis}) and (\ref{gmeanbis2}%
), and for different values of $A_{t}$ at $T=160$ MeV as input, the code
outputs the values of $Y_{B}$ and $Y_{\overline{B}}$ at $T=20$ MeV -- i.e.
at the end of the baryogenesis -- as shown in Fig.~\ref{f1} here for $\kappa
=0.99$ (i.e. $\left\vert \Delta T\right\vert /T\lesssim 10\times 10^{-3})$.
We are obviously focused on $Y_{B}$ when $Y_{\overline{B}}$ is negligible
compared to $Y_{B}$ and with respect with the baryon-antibaryon asymmetry
limit, i.e. when the Universe is dominated by matter (not antimatter) in the
radiative period. Usually, one can assumed that $Y_{\overline{B}%
}/Y_{B}<10^{-6}$ \cite{Canetti2012}. That means that $Y_{B}\sim 8.8\times
10^{-11}$ and $Y_{\overline{B}}<10^{-16}$. Checking for such conditions, as
a striking result, simulations show that the only relevant conditions for
temperatures are

\begin{equation}
7\times 10^{-3}\lesssim \frac{\left\vert \Delta T\right\vert }{T}\lesssim
10\times 10^{-3}.  \label{Tbounds}
\end{equation}%
The upper limit is obviously given by the condition (\ref{cond2}), while the
lower one defines the lowest temperature difference allowing for the
expected baryon abundance and baryon-antibaryon asymmetry. Plots for other
values of $\left\vert \Delta T\right\vert /T$ are not shown as they weakly
differ from the one shown in Fig.~\ref{f1}. Anyway, we can numerically
obtain a continuous function $Y_{B}=Y_{B}(A_{t})$ for a given value of $%
\kappa $. At this step, it is still remarkable to note that our model
constrains a range of relevant values for the magnetic vector potential (see
Fig.~\ref{f1}), if we expect for values of $Y_{B}\sim 8.8\times 10^{-11}$
(see horizontal black dash-dotted line in Fig.~\ref{f1}) in accordance with
observations \cite{Cline2019}. As the blue line, showing $Y_{B}$, cuts the
horizontal black dash-dotted line, it shows that two values of $A_{t}$
allows for the expected baryon abundance assuming that $A_{t}$ is spatially
uniform. These values are shown in Table~\ref{tab:a0_values} for various
conditions on the temperatures. 
\begin{table}[h]
\caption{Lower ($A_{t,\text{low}}$) and higher ($A_{t,\text{high}}$) values
for a spatially uniform magnetic vector potential strength $A_{t}$ required
to produce the observed baryon-antibaryon asymmetry -- and then baryon
density -- against temperature conditions. }
\label{tab:a0_values}\centering{\renewcommand{\arraystretch}{1.5} 
\begin{tabular}{ccc}
\hline\hline
$\left\vert \Delta T\right\vert /T$ & $A_{t,\text{low}}\,(\text{T m}) $ & $%
A_{t,\text{high}}\,(\text{T m})$ \\ \hline
$10\times 10^{-3}$ & $3.3\times 10^{8}$ & $2.6\times 10^{9}$ \\ 
$9\times 10^{-3}$ & $4.0\times 10^{8}$ & $2.4\times 10^{9}$ \\ 
$8\times 10^{-3}$ & $5.1\times 10^{8}$ & $2.2\times 10^{9}$ \\ 
$7\times 10^{-3}$ & $7.3\times 10^{8}$ & $1.8\times 10^{9}$ \\ \hline\hline
\end{tabular}%
}
\end{table}

The behavior of $Y_B$ and $Y_{\bar{B}}$ against $A_t$ reveals a transition
towards an oscillatory regime in the C/CP-violating dynamics. As shown in
Fig.~\ref{f1}, the decrease in baryon density is closely mirrored by a
corresponding increase in the antibaryon density $Y_{\bar{B}}$. This
indicates that the high-field regime induces a shift in the relative
efficiency of matter versus antimatter production during interbrane
transitions. In this context, the specific value of the magnetic field does
not merely scale the asymmetry but determines its sign, leading to
alternating regions of baryon or antibaryon dominance. Consequently, the
observed prevalence of matter in our Universe acts as a selection criterion,
effectively constraining the primordial magnetic field to specific intensity
windows compatible with the observed baryon-to-photon ratio.

This preliminary constraint on the magnetic vector potential already
indicates that the mechanism does not operate efficiently for arbitrary PMF
amplitudes. In particular, only a narrow range of $A_{t}$ values leads to
the observed baryon density, suggesting from the outset that the underlying
primordial magnetic field must lie within a similarly restricted amplitude
range. This anticipates the global consistency requirement on the PMF
strength obtained in Sec. \ref{VA}.

In the following, the local comoving baryon density $Y_{B}(\mathbf{r})$ at $%
T=20$ MeV is modeled as a function of the magnetic vector potential norm $%
A_{t}(\mathbf{r})$ at $T=160$ MeV by $Y_{B}=Y_{B}(A_{t})$. As the matter
density can be fairly approximated by $\rho =msY_{B}$, where $m=939$ MeV/c$%
^{2}$ is the typical baryon mass, we can define the matter over density as 
\begin{equation}
\delta _{b}(\mathbf{r})=\frac{Y_{B}(\mathbf{r})}{\langle Y_{B}\rangle }-1,
\end{equation}%
and we look for the matter power spectrum $P_{\delta }(k)$ defined through%
\begin{equation}
\left\langle \delta _{b}(\mathbf{k})\delta _{b}^{\ast }(\mathbf{k}^{\prime
})\right\rangle =(2\pi )^{3}\delta ^{(3)}(\mathbf{k}-\mathbf{k}^{\prime
})P_{\delta }(k).
\end{equation}

\section{Models of Primordial Magnetic Fields}

\label{III}

Primordial magnetic fields, generated in the early Universe, are key to
cosmological processes, including baryogenesis, structure formation, and CMB
anisotropies \cite{Grasso2001,Subramanian2016,Durrer2013}. Their properties
and upper limits have been extensively constrained by CMB analyses, notably
by the Planck Collaboration \cite{Planck2016}. At the QGP-HG transition, the
spectral properties of this fields are expected to drive the spatial
fluctuations of the magnetic vector potential difference $\mathbf{A}_{+}-%
\mathbf{A}_{-}$, which determines the interbrane coupling in our two-brane
baryogenesis model (see Eq. (\ref{W})) \cite{Sarrazin2024}.

Several mechanisms can generate primordial magnetic fields. During
inflation, magnetogenesis mechanisms can generate large-scale, super-horizon
magnetic fields, with coherence lengths stretched beyond the Hubble radius,
originating from quantum fluctuations of the electromagnetic field \cite%
{Turner1988,Ratra1992}. \textquotedblright Phase
transitions\textquotedblright\ -- or more precisely crossovers in the
Standard Model -- such as the electroweak or QCD transitions \cite%
{Boyanovsky2005,Boeckel2012,Tevzadze2012}, could create \textquotedblright
causal\textquotedblright\ fields \textit{via} bubble collisions or plasma
instabilities \cite{Cheng1994,Vachaspati2021}. We stress that, in the
Standard Model, the QCD transition at vanishing baryon chemical potential is
a smooth crossover rather than a genuine first-order transition. Causal
magnetogenesis at this epoch therefore proceeds primarily through plasma
turbulence and instabilities. Bubble-nucleation or domain-wall dynamics are
relevant only in scenarios in which the QCD transition is rendered first
order -- for instance by a large lepton asymmetry \cite{Boeckel2012} or by
QCD domain walls \cite{Forbes2000}. In what follows, \textquotedblleft
causal\textquotedblright\ refers generically to sub-horizon fields,
irrespective of which of these mechanisms operates, since our analysis
depends only on the resulting power spectrum and not on the details of the
generation process. Post-inflationary turbulence in the primordial plasma
also contributes to fields \cite{Kandus2011,Brandenburg2005}. These
mechanisms define the field's domain structure: super-horizon fields have
coherence lengths exceeding the Hubble horizon, while \textquotedblright
causal\textquotedblright\ fields are limited to sub-horizon scales,
typically $R_{H}\sim cH^{-1}$ ($H$ is the Hubble constant) at QGP-HG
transition \cite{Enqvist1998}. Assuming that the magnetic field is random
and statistically homogeneous, the two-point correlation function in Fourier
space, in the absence of long-range structure or helicity for instance,
takes the form 
\begin{equation}
\left\langle \tilde{B}_{i}(\mathbf{k})\tilde{B}_{j}^{\ast }(\mathbf{k}%
^{\prime })\right\rangle =\delta ^{(3)}(\mathbf{k}-\mathbf{k}^{\prime
})\left( \delta _{ij}-\hat{k}_{i}\hat{k}_{j}\right) P_{B}(k),  \label{corr}
\end{equation}%
where $P_{B}(k)$ is the magnetic power spectrum such that the normalization
of the spectrum is defined from the mean square field, $\sqrt{\langle
B^{2}\rangle }=B_{0}$, i.e. the typical field strength, computed as $\langle
B^{2}\rangle =\frac{1}{(2\pi )^{3}}\int_{0}^{\infty }4\pi k^{2}P_{B}(k)\,dk$%
. We parametrize the PMF power spectrum, $P_{B}(k)$, using a common
phenomenological broken power-law model. This form is chosen to capture the
essential features of magnetic fields generated by turbulent processes in
the early universe, such as those predicted to occur during cosmological
phase transitions \cite{Grasso2001,Kandus2011}. This allows a versatile
framework to study the cosmological implications of such fields while
irrespective of the precise details of the generation mechanism. Then, the
spectrum is defined as \cite{Grasso2001,Kandus2011} 
\begin{equation}
P_{B}(k)=%
\begin{cases}
P_{0}k^{n} & \text{for }k_{\text{IR}}\leq k\leq k_{\ast } \\ 
P_{0}k_{\ast }^{n+m}k^{-m} & \text{for }k_{\ast }<k\leq k_{\text{UV}} \\ 
0 & \text{otherwise}%
\end{cases}%
,  \label{eq:pk_model}
\end{equation}%
with%
\begin{equation}
P_{0}=\text{ }\frac{2\pi ^{2}\left( n+3\right) \left( m-3\right) B_{0}^{2}}{%
\mathcal{N}_{\text{m}}},  \label{Po}
\end{equation}%
where $\mathcal{N}_{\text{m}}$, the denominator of the normalization
constant for the magnetic spectrum, is given by%
\begin{eqnarray}
\mathcal{N}_{\text{m}} &=&\left( m-3\right) \left( k_{\ast
}^{n+3}-k_{IR}^{n+3}\right)  \label{numm} \\
&&+\left( n+3\right) k_{\ast }^{n+m}\left( k_{\ast
}^{-m+3}-k_{UV}^{-m+3}\right) .  \notag
\end{eqnarray}%
The model incorporates sharp cutoffs at an infrared scale $k_{\text{IR}%
}=2\pi H/c$, corresponding to the horizon at the time of magnetogenesis, and
an ultraviolet scale $k_{{UV}}$, representing the dissipation scale where
the turbulent cascade terminates. $k_{\ast }$ is the characteristic
energy-injection scale such that for wave-numbers below $k_{\ast }$, the
spectrum scales as $P_{B}(k)\propto k^{n}$. The characteristic
energy-injection scale $k_{\ast }$ can be deduced from the Alfven velocity 
\cite{Subramanian2016} 
\begin{equation}
v_{A}=\frac{B_{0}c}{\sqrt{\mu _{0}\rho }},  \label{alfven}
\end{equation}%
with $\mu _{0}$ vacuum magnetic permeability, and where%
\begin{equation}
\rho =g_{\ast }(T)\left( \frac{\pi ^{2}}{30}\right) \frac{(k_{B}T)^{4}}{%
(\hbar c)^{3}},  \label{rho}
\end{equation}%
is the energy density during the Big Bang against the Universe temperature $%
T $, with $g_{\ast }(T)$ the associated effective degrees of freedom. Since $%
k_{\ast }\sim 2\pi /L_{\ast }$ with $L_{\ast }\sim v_{A}/H,$ from Eqs. (\ref%
{alfven}) and (\ref{rho}), we deduce 
\begin{equation}
k_{\ast }^{2}\sim \frac{2\pi }{\alpha \,cA_{0}}\sqrt{\mu _{0}\rho }H.
\label{kstarsq}
\end{equation}%
where $A_{0}$ is the magnetic vector potential related to $B_{0}$ through
the equation (\ref{Bo}) demonstrated in the next subsection, and where $%
\alpha $ is a constant defined by Eq. (\ref{alpha}).

The positive spectral index $n>0$ in Eq. (\ref{eq:pk_model}) is a direct
consequence of causality, which suppresses power on super-horizon scales. A
value of $n=2$ (a Batchelor spectrum) is expected for maximally helical
fields, while non-helical fields generated from uncorrelated sources would
lead to $n=4$ \cite{Durrer2003,Caprini2009}. Super-horizon fields generated
during inflation typically produce either a scale-invariant spectrum ($%
P_{B}(k)\propto k^{-3}$) or a blue-tilted spectrum ($n>-3$) \cite{Ratra1992}%
, and are therefore not considered here. Beyond the peak, for $k>k_{\ast }$,
the spectrum becomes a power law $P_{B}(k)\propto k^{-m}$ ($m>0$). This
describes the forward energy cascade within the magnetohydrodynamic (MHD)
inertial range, with $m=11/3$ for Kolmogorov-like turbulence or $m=7/2$ for
Iroshnikov-Kraichnan turbulence \cite{Subramanian2016,Brandenburg2017}.

\subsection{Magnetic Field Behavior During the Cosmic Expansion}

\label{IIIA}

In the temperature range from $160$ MeV to $20$ MeV, corresponding to the
post QGP-HG transition era ($t \sim 10^{-5}$\,s to $t \sim 10^{-2}$\,s), we
assume that the primordial magnetic field influences the spatial
distribution of baryonic matter exclusively through the exotic mechanism
proposed in Ref. \cite{Sarrazin2024}. This mechanism, which involves a
scalar field-induced C/CP symmetry violation in a two-brane universe,
facilitates neutron and antineutron exchanges between the visible and hidden
branes \cite{Sarrazin2024}, altering the baryon-antibaryon asymmetry without
requiring classical MHD interactions. Furthermore, as in Ref. \cite%
{Sarrazin2024}, we assume that the primordial magnetic field $\mathbf{B}$
evolves passively, stretching with the cosmic expansion such that $|\mathbf{B%
}| \propto a^{-2}$. This condition is required to maintain the validity of
the model described in the previous section.

To justify this assumption, we note that the electrical conductivity $\sigma$
of the primordial plasma remains extremely high throughout this epoch \cite%
{Banerjee2004}. Consequently, the magnetic Reynolds number is large ($R_m
\gg 1$), ensuring that the magnetic flux remains frozen into the plasma
(flux freezing limit). Under these conditions, and in the absence of
continuous energy injection or significant forcing to sustain turbulence
after the phase transition, the large-scale magnetic field simply dilutes
with the expansion of the universe \cite{Grasso2001,Banerjee2004}. Complex
MHD effects such as dynamo amplification are therefore not expected to alter
the field strength significantly during this regime.

Furthermore, any back-reaction from baryonic matter onto the magnetic field
is negligible. While the universe is extremely dense, the plasma is
profoundly baryon-poor, characterized by a very small baryon-to-photon ratio
($\eta \sim 10^{-10}$). The number density of baryons is therefore
insufficient to generate electric currents strong enough to alter the
large-scale structure of the primordial field \cite{Banerjee2004}.

As a result, the primordial magnetic field, with a typical strength of $B_0$
at the QCD transition and a coherence length scaling as $L \propto a$,
evolves almost exclusively through the passive stretching of its field lines
by cosmic expansion ($|\mathbf{B}| \propto a^{-2}$) \cite%
{Cheng1994,Subramanian2016,Brandenburg2005}, maintaining a constant spectral
shape in comoving coordinates \cite{Kahniashvili2016}. We may also note that
from Eq. (\ref{kstarsq}) and from $k_{\text{IR}}$ we get 
\begin{eqnarray}
\frac{k_{\ast }}{k_{\text{IR}}} &=&\frac{\left( \mu _{0}\rho \right) ^{1/4}}{%
\left( \alpha \,A_{0}k_{\text{IR}}\right) ^{1/2}}  \label{kratio} \\
&=&\frac{c}{\left( \alpha \,A_{0}\right) ^{1/2}}\left( \frac{3\mu _{0}}{%
32\pi ^{3}G_{N}}\right) ^{1/4},  \notag
\end{eqnarray}
meaning that the ratio $k_*/k_{\text{IR}}$ is constant during the early
evolution of the Universe. Thus, both complex MHD interactions and
significant matter feedback are negligible, ensuring that the spatial
statistics of baryonic matter are modulated solely by the interbrane
coupling mechanism, without altering the magnetic field spectrum. This
approximation holds until temperatures drop below $20$ MeV, where
recombination and subsequent processes may introduce additional dynamics 
\cite{Durrer2013}. However, as shown in Ref. \cite{Sarrazin2024}, the
baryonic density and the baryon-antibaryon asymmetry are frozen out well
before $20$ MeV.

\subsection{Spectrum of the Magnetic Potential}

\label{IIIB}

As explained above, the spectrum of $A_{t}=\left\vert \mathbf{A}%
_{t}\right\vert =|\mathbf{A}_{+}-\mathbf{A}_{-}|$ drives the fluctuations of 
$Y_{B}$. The magnetic fields $\mathbf{B}_{\pm }=\nabla \times \mathbf{A}%
_{\pm }$ on each brane derive from the vector potentials $\mathbf{A}_{\pm }$%
. In Fourier space, assuming the Coulomb gauge ($\nabla \cdot \mathbf{A}%
_{\pm }=0$) and Gaussian statistics, the power spectrum of $\mathbf{A}_{\pm
} $, $P_{A}(k)$, relates to $P_{B}(k)$ \textit{via} 
\begin{equation}
P_{A}(k)=\frac{P_{B}(k)}{k^{2}},  \label{PA}
\end{equation}%
since $\mathbf{B}(\mathbf{k})=i\mathbf{k}\times \mathbf{A}(\mathbf{k})$ \cite%
{Subramanian2016}. This spectral behavior reflects the fact that the vector
potential is smoother than the field itself, due to the suppression of
small-scale fluctuations by the $1/k^{2}$ factor. This gauge choice is not
merely a computational convenience. Under a gauge transformation $\mathbf{A}%
_{\pm }\rightarrow \mathbf{A}_{\pm }+\nabla \Lambda _{\pm }$ the added
gradient is purely longitudinal ($i\mathbf{k}\,\Lambda _{\pm }(\mathbf{k})$
in Fourier space), so that the transverse component retained by the Coulomb
condition $\nabla \cdot \mathbf{A}_{\pm }=0$ is strictly gauge invariant: it
carries the physical, transverse degrees of freedom of the field. The local
amplitude $A_{t}(x)$, the rate $\varepsilon $ of Eq.~(\ref{W}), and all
statistics derived from them are therefore genuine gauge-invariant
quantities, fixed by the gauge-invariant magnetic spectrum $P_{B}(k)$. This
complements, on the side of the field amplitude, the invariance of the
transition energy discussed in Sec.~\ref{IIA}: both the magnitude and the
phase of $\mathbf{A}_{+}-\mathbf{A}_{-}$ enter the observables only through
gauge-invariant combinations. Then, from Eqs. (\ref{PA}) and (\ref%
{eq:pk_model}), we get

\begin{equation}
P_{A}(k)=%
\begin{cases}
P_{0}k^{n-2} & \text{for }k_{\text{IR}}\leq k\leq k_{\ast } \\ 
P_{0}k_{\ast }^{n+m}k^{-m-2} & \text{for }k_{\ast }<k\leq k_{\text{UV}} \\ 
0 & \text{otherwise}%
\end{cases}%
,  \label{PAdet}
\end{equation}%
with now 
\begin{equation}
P_{0}=\frac{2\pi ^{2}\left( n+1\right) \left( m-1\right) A_{0}^{2}\text{ }}{%
\mathcal{N}_{\text{p}}},  \label{PoA}
\end{equation}%
where $\mathcal{N}_{\text{p}}$, the denominator of the normalization
constant for the spectrum of the magnetic potential, is given by 
\begin{eqnarray}
\mathcal{N}_{\text{p}} &=&\left( m-1\right) \left( k_{\ast
}^{n+1}-k_{IR}^{n+1}\right)  \label{nump} \\
&&+\left( n+1\right) k_{\ast }^{n+m}\left( k_{\ast
}^{-m+1}-k_{UV}^{-m+1}\right) ,  \notag
\end{eqnarray}%
such that the normalization of the spectrum is defined from the mean square
field, $\sqrt{\langle A^{2}\rangle }=A_{0}$, i.e. the typical field
strength, computed as $\langle A^{2}\rangle =\frac{1}{(2\pi )^{3}}%
\int_{0}^{\infty }4\pi k^{2}P_{A_{t}}(k)\,dk$. From Eqs. (\ref{Po}) and (\ref%
{PoA}), we deduce

\begin{eqnarray}
B_{0} &=&A_{0}\sqrt{\frac{\left( n+1\right) \left( m-1\right) }{\left(
n+3\right) \left( m-3\right) }}\sqrt{\frac{\mathcal{N}_{\text{m}}}{\mathcal{N%
}_{\text{p}}}}  \label{Bo} \\
&\sim &\alpha \,A_{0}\,k_{\ast }\text{ if }k_{\text{IR}}\ll k_{\ast }\ll k_{%
\text{UV}},  \notag
\end{eqnarray}%
with%
\begin{equation}
\alpha \,=\sqrt{\frac{\left( n+1\right) \left( m-1\right) }{\left(
n+3\right) \left( m-3\right) }}.  \label{alpha}
\end{equation}%
Considering now $A_{t}=\left\vert \mathbf{A}_{t}\right\vert =|\mathbf{A}_{+}-%
\mathbf{A}_{-}|$ the related power spectrum is 
\begin{equation}
P_{A_{t}}(k)=\langle |\mathbf{A}_{+}-\mathbf{A}_{-}|^{2}\rangle
_{k}=P_{A}(k)[2-2\rho (k)],  \label{Padeff}
\end{equation}%
where $\rho (k)=\langle \mathbf{A}_{+}(\mathbf{k})\cdot \mathbf{A}_{-}^{\ast
}(\mathbf{k})\rangle /[P_{A}(k)]^{1/2}$ is the cross-correlation
coefficient, and we assume identical $P_{A}(k)$ for both branes at QGP-HG
transition and after. Partial correlation ($0<\rho (k)<1$) may arise from
interbrane interactions, which can influence the electromagnetic fields on
each brane. In the current version of our model \cite%
{Sarrazin2024,Sarrazin2010,Stasser2019}, such effects are either absent or
negligible, so we assume $\rho (k)\approx 0$ in the following. The resulting
statistics will be used in Sec. \ref{methodology} to compute the spatial
distribution of the baryon asymmetry.

\subsection{One-Point Statistics of the Magnetic Vector Potential Amplitude}

\label{IIIC}

Beyond its two-point statistics encoded in the power spectrum $P_{A_{t}}(k)$%
, the baryogenesis mechanism introduced in Sec. \ref{II} depends locally on
the magnitude of the magnetic vector potential difference, $A_{t}(x)=|%
\mathbf{A}_{+}(x)-\mathbf{A}_{-}(x)|$. It is therefore essential to specify
the one-point probability distribution of $A_{t}$.

Under the assumptions adopted throughout this work, the magnetic vector
potentials on each brane, $\mathbf{A}_{\pm}$, are modeled as statistically
homogeneous and isotropic Gaussian random fields. This hypothesis is
standard in the literature on primordial magnetic fields and applies both to
inflationary and causal magnetogenesis scenarios, at least at the level of
the vector potential \cite{Grasso2001,Subramanian2016,Durrer2013}. This
implies that, at any given spatial point, the Cartesian components $A_{\pm
,i}$ $(i=1,2,3)$ are independent Gaussian random variables with identical
variance, 
\begin{equation}
\langle A_{\pm ,i}\rangle =0,\qquad \langle A_{\pm ,i}A_{\pm ,j}\rangle
=\sigma _{A}^{2}\delta _{ij}.  \label{stat}
\end{equation}%
with $\sigma _{A}=A_{0}/\sqrt{3}$, such that $\sigma_{A}^2$ is the variance
of the field components on a single brane.

Assuming negligible cross-correlation between the branes (as discussed in
Sec. \ref{IIIB}), the difference vector field $\mathbf{A}_{t}=\mathbf{A}_{+}-%
\mathbf{A}_{-}$ is itself an isotropic Gaussian random field. Its Cartesian
components $A_{t,i}$ ($i=1,2,3$) are therefore independent Gaussian random
variables with zero mean and a total variance given by the sum of the
individual variances: 
\begin{equation}
\langle A_{t,i}A_{t,j}\rangle =\sigma _{A_{t}}^{2}\delta _{ij}\quad \text{%
with}\quad \sigma _{A_{t}}^{2}=2\sigma _{A}^{2}.  \label{OP}
\end{equation}

As a direct and model-independent consequence, the amplitude $A_{t}=|\mathbf{%
A}_{t}|=\sqrt{A_{t,x}^{2}+A_{t,y}^{2}+A_{t,z}^{2}}$ follows a Maxwellian
distribution defined by the parameter $\sigma _{A_{t}}$: 
\begin{equation}
f(A_{t})dA_{t}=\sqrt{\frac{2}{\pi }}\frac{A_{t}^{2}}{\sigma _{A_{t}}^{3}}%
\exp \left( -\frac{A_{t}^{2}}{2\sigma _{A_{t}}^{2}}\right) dA_{t}.
\label{eq:maxwell_At}
\end{equation}%
This result arises purely from the Gaussian nature of the vector field
components and rotational invariance, independently of the detailed shape of
its power spectrum \cite{Adler1981}. Consequently, when computing the mean
baryon density $\langle Y_{B}\rangle $ induced by the two-brane baryogenesis
mechanism, the statistical averaging over realizations of the magnetic field
is consistently performed using the distribution (\ref{eq:maxwell_At}) for
the variable $A_{t}$. This prescription is independent of the specific
magnetogenesis mechanism, of the spectral indices $(n,m)$, and of the
spatial correlation properties discussed in the following section, and
constitutes a general statistical foundation of our semi-analytical approach.

\section{Semi-Analytical Computation of the Baryonic Density Fluctuation
Power Spectrum}

\label{methodology}

We use a semi-analytical method to compute the power spectrum of baryonic
density fluctuations, $P_{\delta }(k)$, induced by a primordial magnetic
vector potential. Our approach follows the classical Wiener-Khinchin
framework for relating correlation functions and power spectra in cosmology 
\cite{Peebles1980,BBKS1986}, and is similar in spirit to the semi-analytical
treatments of stochastic magnetic fields developed in \cite%
{Caprini2001,Durrer2003,Caprini2009}. This method avoids the computational
cost of full three-dimensional grid simulations, while retaining sufficient
accuracy to capture the relevant physical processes.

The calculation proceeds in three main steps: (1) computing the
auto-correlation function $C(r)$ of the magnetic vector potential from its
power spectrum $P_{A_{t}}(k)$; (2) estimating the auto-correlation of
baryonic density fluctuations $\xi _{\delta }(r)$ through a two-point Monte
Carlo sampling of correlated Gaussian random fields, following the general
statistical approaches to Gaussian random fields \cite{Adler1981,Coles1991};
and (3) obtaining $P_{\delta }(k)$ by applying a Fourier transform to $%
\xi_{\delta }(r)$. The Gaussian assumption is consistent with the standard
treatment of PMF as a Gaussian stochastic field, with non-Gaussianity
arising only at the level of anisotropies due to the quadratic nature of the
energy-momentum tensor \cite{Planck2016}. The details of the numerical
implementation are presented below and the related \texttt{PrimoSpec.py}
code is provided in the ''Data and Code Availability'' section of the
present paper.

\subsection{Auto-Correlation of the Magnetic Vector Potential}

The magnetic vector potential $\mathbf{A}$ is modeled as a Gaussian random
field with an isotropic power spectrum $P_{A_{t}}(k)$, defined by Eq. (\ref%
{PAdet}). The two-point auto-correlation $C(r)=\left\langle \mathbf{A}_{t}%
\mathbf{(x)\cdot A}_{t}\mathbf{(x+r)}\right\rangle $ is obtained \textit{via}
the Fourier transform 
\begin{equation}
C(r)=\frac{1}{2\pi ^{2}}\int_{k_{\text{IR}}}^{k_{\text{UV}}}k^{2}P_{A_{t}}(k)%
\frac{\sin (kr)}{kr}\,dk.  \label{eq:Cr}
\end{equation}%
Numerically, this integral is evaluated over a logarithmic grid of $N_{k}$
wave-numbers from $k_{\text{IR}}$ to $k_{\text{UV}}$, using trapezoidal
integration for each of $N_{r}$ logarithmically spaced distances $r$ ranging
from $r_{\min }=0.5/k_{\text{UV}}$ to $r_{\max }=2R_{H}$. The resulting $%
C(r) $ is interpolated cubically to ensure smooth evaluation in subsequent
steps.

\subsection{Auto-Correlation of the Baryonic Density Fluctuations}

The comoving baryon density $Y_{B}$ at $T=20$ MeV can be modeled as a
function of the magnetic vector potential norm $A_{t}$ based on empirical
fits from the prior simulations \cite{Sarrazin2024} shown in Sec. \ref{IIC}.
The plot in Fig.~\ref{f1} shows a complex pattern, for which only a
non-trivial parametrization achieves the required numerical accuracy. In the
interval $I_{A}=$ $\left[ A_{t,\text{low}};A_{t,\text{high}}\right] $ (see
Table~\ref{tab:a0_values}), $Y_{B}(A_{t})$ exhibits significantly higher
values, by several orders of magnitude, in contrast to those outside of $%
I_{A}$, and when compared to $Y_{\overline{B}}(A_{t})$. In addition to the
Maxwellian distribution (\ref{eq:maxwell_At}) of $A_{t}$, the vicinity of $%
I_{A}$ is the domain of interest for$\ Y_{B}(A_{t})$, which can be modeled
within this restricted range of values. Then, we numerically obtain a
continuous function $Y_{B}=Y_{B}(A_{t})$ defined as 
\begin{equation}
Y_{B}(A_{t})=Y_{0}\left( 2.61-\chi \right) \chi ^{1.58-\chi ^{1/3}},
\label{fit}
\end{equation}%
where $\chi =\left( A_{t}-A_{t,0}\right) /S$ with $Y_{0}=1.07\times 10^{-10}$%
, $A_{t,0}=1.01\times 10^{6}$ Tm and $S=1.25\times 10^{9}$ Tm, where --
without loss of generality -- we have considered $\kappa =0.991$, which is
used throughout the remainder of this work. To obtain this analytical
representation of the numerical results presented in Fig.~\ref{f1}, we
performed a symbolic regression using the \textit{PySR} library based on the 
\textit{PySRRegressor} class. This approach allowed us to derive the
functional form of $Y_{B}(A_{t})$ defined in Eq.~(\ref{fit}) with high
accuracy. In addition, for values $A_{t}<A_{t,0}$, we set $Y_{B}=10^{-19}$,
which corresponds to the expected comoving density without baryogenesis
mechanism \cite{Sarrazin2024}. That means for weak fields ($A_{t}<A_{t,0}$),
the C/CP violation is insufficient to allow the baryogenesis. This
approximation leads to no substantial error to the full computation.

Then, density fluctuations are defined as \newline
$\delta = Y_{B}(A_{t})/\langle Y_{B}\rangle -1$, where the mean density $%
\langle Y_{B}\rangle $ is computed thanks to the Maxwellian distribution (%
\ref{eq:maxwell_At}) of $A_{t}$, with variance $\sigma _{A_{t}}^{2}$ derived
from Step 1. The auto-correlation $\xi _{\delta }(r)=\left\langle \delta 
\mathbf{(x)}\delta \mathbf{(x+r)}\right\rangle $ is estimated using a
two-point Monte Carlo method. For each distance $r$, we generate $N_{\text{%
pairs}}$ pairs of correlated Gaussian vectors $\mathbf{A}_{1}$ and $\mathbf{A%
}_{2}$, with correlation coefficient $\rho (r)=C(r)/\sigma _{A_{t}}^{2}$
(clipped to $[-1,1]$) 
\begin{equation}
\mathbf{A}_{1}=\sigma _{A_{t}}\mathbf{Z}_{1},\quad \mathbf{A}_{2}=\sigma
_{A_{t}}\left( \rho \mathbf{Z}_{1}+\sqrt{1-\rho ^{2}}\mathbf{Z}_{2}\right) ,
\label{eq:MC_vectors}
\end{equation}%
where $\mathbf{Z}_{1,2}$ are standard normal vectors. In practice, the
Maxwellian statistics of the vector potential amplitude arises naturally
from sampling a three-dimensional isotropic Gaussian random vector and
computing its norm. The auto-correlation $\xi _{\delta }(r)$ is then
computed as the sample mean of $\delta _{1}\delta _{2}$. This approach
efficiently captures the non-linearity of $Y_{B}(A_{t})$ without requiring
full-field simulations. The resulting $\xi _{\delta }(r)$ is linearly
interpolated for use in the next step.

\subsection{Power Spectrum \textit{via} Fourier Transform}

The power spectrum $P_{\delta }(k)$ is obtained \textit{via} the Fourier
transform of $\xi _{\delta }(r)$ 
\begin{equation}
P_{\delta }(k)=4\pi \int_{0}^{r_{\max }}r^{2}\xi _{\delta }(r)\frac{\sin (kr)%
}{kr}\,dr.  \label{eq:Pdelta}
\end{equation}%
with an integration cutoff $r_{\max }$ optimized for convergence and defined
by the value for which $\xi _{\delta }(r)$ falls to zero.

\subsection{Numerical Implementation and Validation}

All computations are implemented in Python 3.10, using \textit{NumPy} for
array operations, \textit{SciPy} for numerical integration and
interpolation. The Monte Carlo simulation in Step 2 dominates the runtime
(on the order of minutes on standard hardware) but scales linearly with $N_{%
\text{pairs}}$, allowing precision tuning (achieving variance $\sim 10^{-6}$%
). Integration accuracy is monitored through quadrature errors, typically
below $10^{-10}$. The use of logarithmic grids ensures resolution across
multiple scales, while direct integration avoids fast Fourier transform
artifacts. Validation against analytical limits (e.g., white-noise regimes)
confirms physical consistency (see Sec. \ref{VB}). This semi-analytical
framework provides a robust and memory-efficient alternative to grid-based
simulations, making it well-suited for exploring parameter spaces in
primordial cosmology. The method's flexibility and scalability facilitate
further extensions, such as incorporating additional physical processes or
refining numerical precision.

\section{Results and Interpretation}

\label{sec:results}

\subsection{Global Baryon Density: Magnetic Field Amplitude Required}

\label{VA}

The restricted range of magnetic-vector-potential amplitudes identified in
Sec. \ref{IIC} already hinted that the baryogenesis mechanism is predictive
rather than merely permissive. We now demonstrate explicitly that this local
constraint on $A_t$ translates into a global requirement on the primordial
magnetic-field strength $B_0$.

A crucial first step is to verify that our baryogenesis mechanism, when
seeded by a stochastic magnetic field, can reproduce the observed global
baryon density, $Y_{B,\text{obs}}=(8.8\pm 0.6)\times 10^{-11}$ \cite%
{Cline2019}. We integrate the baryon density $Y_{B}\left( A_{t}\right) $
generated over the statistical distribution $f(A_{t})$ (see Eq. \ref%
{eq:maxwell_At}) for various values of $A_{0}$ of the magnetic vector
potential in order to get the average baryon density $\langle Y_{B}\rangle
=Y_{B,\text{obs}}$ in the visible universe, and then to determine the
required typical magnetic field strength, $B_{0}=\sqrt{\langle B^{2}\rangle }
$ (using Eq. (\ref{Bo})), at the QGP-HG transition. The results, in Table %
\ref{tab:b0_values}, show that for the different \textquotedblright
causal\textquotedblright\ spectra under study (varying spectral index $n$),
the required field strength is consistently around $10^{10}$ T \cite%
{Brandenburg1996,Forbes2000,Boyanovsky2005,Boeckel2012,Tevzadze2012}. This
value aligns remarkably well with theoretical predictions for magnetic
fields generated during the QCD phase transition \cite%
{Brandenburg1996,Forbes2000}, providing a strong consistency check for the
model. We note that the plot in Fig.~\ref{f1}, which illustrates the
non-linear nature of the baryogenesis efficiency, suggests that, for each
spectral index $n$, two possible values of $A_{0}$, and consequently of $%
B_{0}$, can produce the correct baryon density (see Table~\ref{tab:a0_values}%
). But as the distribution $f(A_{t})$ of $A_{t}$ spreads over a wide range
of values, $A_{t}$ can reach magnitudes that allow for an antibaryon density
exceeding the observed value. Only smaller values of $A_{t}$ (see Table~\ref%
{tab:a0_values}), and consequently of $A_{0}$ and $B_{0}$, allow for the
correct baryon and antibaryon densities.

\begin{table}[th]
\caption{Strength of the magnetic field $B_{0}$ (T) required to produce the
observed baryon-antibaryon asymmetry and then baryon density, for different
PMF spectral indices $n$ and relative difference of temperature $\left\vert
\Delta T\right\vert /T$. }
\label{tab:b0_values}\centering{\renewcommand{\arraystretch}{1.5} 
\begin{tabular}{cccc}
\hline\hline
$\left\vert \Delta T\right\vert / T$ & $n=0$ & $n=2$ & $n=4$ \\ \hline
$10 \times 10^{-3}$ & $6.6\times 10^{9}$ & $7.7\times 10^{9}$ & $8.0\times
10^{9}$ \\ 
$9 \times 10^{-3}$ & $7.3\times 10^{9}$ & $8.5\times 10^{9}$ & $8.9\times
10^{9}$ \\ 
$8 \times 10^{-3}$ & $8.4\times 10^{9}$ & $9.8\times 10^{9}$ & $1.0\times
10^{10}$ \\ 
$7 \times 10^{-3}$ & $1.1\times 10^{10}$ & $1.3\times 10^{10}$ & $1.3\times
10^{10}$ \\ \hline\hline
\end{tabular}%
}
\end{table}

In this sense, the constraint on the magnetic potential derived in Sec. \ref%
{IIC} finds its physical realization here: only those PMF amplitudes that
reproduce the required range of $A_t$ simultaneously yield the correct
global baryon density. This demonstrates the consistency between the
microscopic interbrane dynamics and the macroscopic PMF properties.

It is crucial to emphasize the highly non-trivial nature of this agreement.
The baryogenesis mechanism we propose, based on brane physics and C/CP
violation induced by a pseudo-scalar field \cite{Sarrazin2024}, is \textit{a
priori} entirely disconnected from magnetogenesis mechanisms stemming from
plasma dynamics during the QCD transition \cite%
{Brandenburg1996,Forbes2000,Boyanovsky2005,Boeckel2012,Tevzadze2012}.

The fact that our model, in order to explain the observed baryon asymmetry,
robustly selects a PMF amplitude that precisely coincides with that expected
from hydrodynamic considerations (e.g., turbulence and plasma instabilities)
is not a mere compatibility check. It suggests a deep convergence between
the beyond-Standard-Model physics (required for baryogenesis) and the
Standard Model physics (the QGP-HG transition) at this epoch. The PMF
amplitude is no longer a posited free parameter, but rather a predictive
consequence linking these two domains.

In addition, the required field strength is moreover remarkably insensitive
to the assumed spectrum: across the full range of spectral indices $n=0,2,4$
and brane temperature mismatches considered (Table~\ref{tab:b0_values}), $%
B_{0}$ varies by less than a factor of two. This follows directly from Eq.~(%
\ref{Bo}): since $B_{0}\sim \alpha \,A_{0}\,k_{\ast }$ with $\alpha =%
\mathcal{O}(1)$ depending only weakly on $(n,m)$, the amplitude is
controlled by the injection scale $k_{\ast }$ rather than by the infrared or
ultraviolet cutoffs. The latter enter only through the spectral
normalization, whose integrals converge near $k_{\ast }$ in the regime $%
k_{IR}\ll k_{\ast }\ll k_{UV}$. Note that the white-noise amplitude $%
P_{\delta ,0}$ (Table~\ref{tab:p0_values}) -- discussed in the next section
-- is more sensitive to $n$, but lies many orders of magnitude below the CMB
bound for all cases (see Sec.~\ref{VC}), so this residual sensitivity has no
observational consequence.

\subsection{Baryon-Density Fluctuation Spectrum: Universal White-Noise
Behavior}

\label{VB}

\begin{figure}[ht]
\centering
\includegraphics[width=8.5cm]{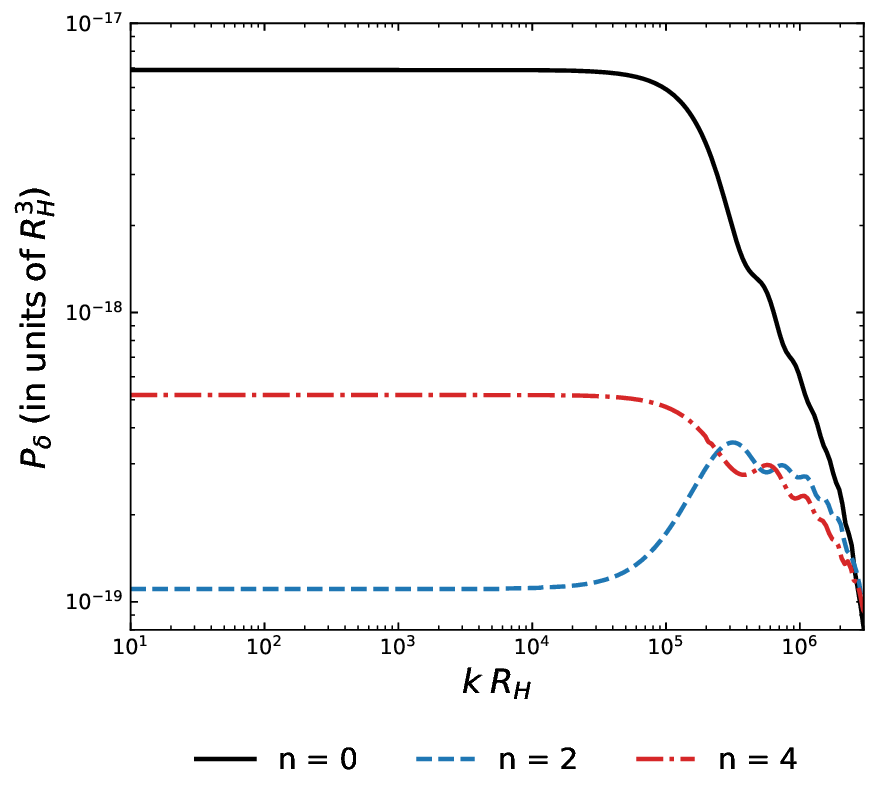}
\caption{The power spectrum of baryon density fluctuations, $P_\protect\delta%
(k)$, at $T=20 \, \text{MeV}$ and for $\protect\kappa = 0.991$ ($\left\vert
\Delta T\right\vert /T = 9 \times 10^{-3}$, see Table~\protect\ref%
{tab:b0_values}) as a function of the comoving wave-number $k$. The
different curves correspond to PMF models with different spectral indices $n$
and the magnetic field strengths $B_{0}$ tuned to produce the correct mean
baryon density. Regardless of the input PMF spectrum, the output is a
white-noise spectrum ($P_\protect\delta(k) \approx \text{const}$) for $k <
k_*$.}
\label{f2}
\end{figure}

We now established the model's viability regarding the power spectrum of the
induced baryon density fluctuations, $P_{\delta }(k)$. Using the
semi-analytical method described in Sec. \ref{methodology}, we compute $%
P_{\delta }(k)$ for each of the PMF models that satisfy the global baryon
density constraint. Without loss of generality, we choose to show the case
such that $\kappa =0.991$ ($\left\vert \Delta T\right\vert /T=9\times
10^{-3} $, see Table~\ref{tab:b0_values}). Other situations are not shown,
but do not differ significantly. The results, plotted in Fig.~\ref{f2},
reveal a striking and robust feature: for all considered PMF spectra ($n\geq
0$), the resulting baryon fluctuation spectrum is a pure white noise ($%
P_{\delta }(k)\propto k^{0}$) up to the PMF characteristic scale $k_{\ast }$%
. As a result, white-noise amplitudes, $P_{\delta ,0}=\lim_{k\rightarrow
0}P_{\delta }(k)$, are listed in Table~\ref{tab:p0_values}.

A fundamental explanation of this universality follows directly from the
structure of the correlation functions involved. As defined above, the
baryon-density contrast is given by $\delta (\mathbf{x})=Y_{B}(A_{t}(\mathbf{%
x}))/\langle Y_{B}\rangle -1$, with two-point correlation function $\xi
_{\delta }(r)=\langle \delta (\mathbf{x})\delta (\mathbf{x}+\mathbf{r}
)\rangle $. As shown in Sec. \ref{methodology}, the baryon power spectrum is
obtained through the spherical Fourier transform 
\begin{equation}
P_{\delta }(k)=4\pi \!\int_{0}^{\infty }\!r^{2}\,\xi _{\delta }(r)\,\frac{%
\sin (kr)}{kr}\,dr.  \label{Pdelta_univ}
\end{equation}%
For any PMF spectrum considered in Sec. \ref{III}, the magnetic vector
potential $A_{t}$ has a finite variance and a finite correlation length $%
L_{c}\sim k_{\ast }^{-1}$, implying that $\xi _{\delta }(r)$ decays rapidly
for $r\gtrsim L_{c}$. Because $Y_{B}(A_{t})$ is a local and strongly
non-linear mapping, $\xi _{\delta }(r)$ inherits this finite support and
does not depend on the detailed power-law index $n$ at scales $k\ll k_{\ast
} $.

In the long-wavelength limit, the kernel obeys \newline
$sin(kr)/(kr)\rightarrow 1$ as $k\rightarrow 0$, and Eq.~(\ref{Pdelta_univ})
reduces to 
\begin{equation}
P_{\delta }(0)=4\pi \!\int_{0}^{\infty }\!r^{2}\,\xi _{\delta
}(r)\,dr=P_{\delta ,0}=\mathrm{const.}  \label{WhiteNoiseConst}
\end{equation}%
The integral converges because $\xi _{\delta }(r)$ vanishes for $r\gtrsim
L_{c}$, and the constant depends only on local moments of the joint
distribution of $A_{t}$, together with the non-linear mapping $Y_{B}(A_{t})$%
, but not on the PMF spectral index $n$. Consequently, any short-range
correlated Gaussian field subjected to a sufficiently local non-linear map
inevitably produces a white-noise spectrum at scales $k\ll k_{\ast }$,
explaining the universal flat behavior observed in Fig.~\ref{f2}. It is
important to emphasise that the emergence of a white-noise baryon spectrum
at $k\ll k_{\ast }$ does not rely on the specific broken power-law PMF model
introduced in Sec. \ref{III}. The result is completely generic for any
Gaussian primordial magnetic field. Causality implies that the magnetic
potential $A_{t}$ is a smooth Gaussian field with a finite correlation
length set by the horizon or by the injection scale $k_{\ast }^{-1}$, so
that its two-point function decays rapidly for $r\gtrsim L_{c}$. Since the
baryon asymmetry $Y_{B}(A_{t})$ is a local and strongly non-linear function
of $A_{t}$, the induced baryon density contrast inherits this finite-range
structure. Consequently, the correlation function $\xi _{\delta }(r)$ has
compact (or effectively compact) support, and its Fourier transform reduces
to a constant in the limit $k\rightarrow 0$, independently of the detailed
values of $(n,m)$ or of the precise shape of the PMF spectrum. The
white-noise behaviour is therefore a robust, model-independent prediction of
the mechanism, following solely from short-range correlated Gaussian
statistics combined with a local non-linear mapping. 
\begin{table}[th]
\caption{The white-noise amplitude $P_{\protect\delta ,0}$ of the baryon
fluctuation power spectrum for various PMF models at $T=20$ MeV and for $%
\protect\kappa =0.991$.}
\label{tab:p0_values}\centering{\renewcommand{\arraystretch}{1.5} 
\begin{tabular}{cc}
\hline\hline
Spectral index $n$ & $P_{\delta ,0}\,(\text{R}_{H}^{3})$ \\ \hline
$0$ & $6.7\times 10^{-18}$ \\ 
$2$ & $1.1\times 10^{-19}$ \\ 
$4$ & $5.2\times 10^{-19}$ \\ \hline\hline
\end{tabular}%
}
\end{table}
The white-noise behavior derived above fully characterizes the statistical
properties of the baryon fluctuations produced by our mechanism. Now, it is
useful to clarify which features of these results are model independent and
which are specific to the present scenario. The shape of the spectrum at $%
k\ll k_{\ast }$ -- its flat, white-noise form -- is generic: as shown above,
it follows solely from the short-range correlations of a causal Gaussian
magnetic potential combined with a local non-linear mapping, and is
insensitive to the spectral indices $(n,m)$, to the precise PMF generation
mechanism, and to the details of the brane dynamics. Far from being a
weakness, this universality is what makes the qualitative prediction robust
against the poorly known microphysics of QCD-era magnetogenesis. The
location of the spectral break $k_{\ast }$ is set by the PMF injection scale
(see Eq.~(\ref{kstarsq})) and is therefore a property of the magnetogenesis
model, not of the baryogenesis mechanism. By contrast, two features are
genuinely specific to the two-brane scenario. First, the white-noise
amplitude $P_{\delta ,0}$ (Table~\ref{tab:p0_values}) is fixed by the
non-linear transfer function $Y_{B}(A_{t})$ (see Eq.~(\ref{fit})) and by the
one-point statistics of $A_{t}$; it encodes the brane-specific
C/CP-violating dynamics and could not be inferred from the PMF statistics
alone. Second, and most importantly, the resulting fluctuations constitute a
pure, statistically independent baryon-isocurvature mode -- as discussed in
the next section \ref{VC}: this decoupling from the adiabatic mode is a
non-trivial consequence of the underlying interbrane Boltzmann dynamics and
is not a generic property of non-linearly processed Gaussian fields. In
summary, the mechanism predicts a white-noise shape generically, but fixes
its amplitude and its isocurvature, uncorrelated character through
model-specific physics. Before turning to observational
considerations, it is then important to clarify the physical nature of these
fluctuations and their relation to the pre-existing adiabatic perturbations.
This is the purpose of the next subsection.

\subsection{Subdominant Baryon--Isocurvature Component Compatibility With
CMB Constraints}

\label{VC}

The next critical step is to understand the nature of these newly generated
fluctuations within the standard cosmological framework. The baryon density
contrast, $\delta _{b}(\mathbf{r})=(Y_{B}(\mathbf{r})-\langle Y_{B}\rangle
)/\langle Y_{B}\rangle $, is sourced by our mechanism. A key question is how
it relates to the pre-existing adiabatic fluctuations, $\delta _{adia}$,
inherited from inflation, for instance. To clarify this issue, let us
considers the additional presence of pre-existing fluctuations in the
initial baryon-antibaryon density field, $Y_{B\overline{B}}(\mathbf{r})$.
These fluctuations are described by a density contrast $\delta _{B\overline{B%
}}(\mathbf{r})=(Y_{B\overline{B}}(\mathbf{r})-\left\langle Y_{B\overline{B}%
}\right\rangle )/\left\langle Y_{B\overline{B}}\right\rangle $ around the
mean value $\left\langle Y_{B\overline{B}}\right\rangle $ -- given by the
Boltzmann distribution -- and are characterized by an initial power spectrum 
$P_{0}(k)$. The baryogenesis mechanism now depends on both fields, such that
the final baryon density is given by a function $Y_{B}^{\ast }(\mathbf{r}%
)=Y_{B}(A_{t}(\mathbf{r}),Y_{B\overline{B}}(\mathbf{r}))$. Assuming small
fluctuations, we can perform a first-order expansion of the resulting baryon
density contrast \cite{Peebles1980}, $\delta (\mathbf{r})=(Y_{B}^{\ast }(%
\mathbf{r})-\left\langle Y_{B}^{\ast }\right\rangle )/\left\langle
Y_{B}^{\ast }\right\rangle $, around the mean value $\left\langle
Y_{B}^{\ast }\right\rangle $ such that 
\begin{equation}
\delta (\mathbf{r})\approx \delta _{b}(\mathbf{r})+C\,\delta _{B\overline{B}%
}(\mathbf{r}),  \label{Lincomb}
\end{equation}%
where $\delta _{b}$ is the fluctuation sourced by the magnetic field alone,
and the coefficient $C$ represents the efficiency of the baryogenesis
process in transferring initial density fluctuations to the final baryon
density field. It is defined as 
\begin{equation}
C=\frac{\left\langle Y_{B\overline{B}}\right\rangle }{\left\langle
Y_{B}^{\ast }\right\rangle }\left. \frac{\partial Y_{B}(A_{t},Y_{B\overline{B%
}})}{\partial Y_{B\overline{B}}}\right\vert _{A_{t}=\left\langle
A_{t}\right\rangle ,Y_{B\overline{B}}=\left\langle Y_{B\overline{B}%
}\right\rangle }.  \label{C}
\end{equation}%
A remarkable feature of our model is that, at linear order, it is entirely
decoupled from the initial adiabatic density field. Indeed, a numerical
calculation based on the underlying Boltzmann equations (detailed in Ref. 
\cite{Sarrazin2024}) shows that 
\begin{equation}
\frac{\partial Y_{B}^{\ast }}{\partial Y_{B\overline{B}}}\bigg|_{\langle
A_{t}\rangle ,\langle Y_{B\overline{B}}\rangle }=0,  \label{eq:decoupling}
\end{equation}%
such that the first-order transfer coefficient $C$ is identically zero. Note
that higher-order contributions \cite{Peebles1980} are suppressed by powers
of the initial density contrast $\delta _{B\overline{B}}(r)$. As this
contrast is the imprint of the primordial adiabatic mode, its amplitude is
well-known: $\delta _{B\overline{B}}=\delta _{\text{adia}}\sim 10^{-5}$, at
the baryogenesis epoch \cite{Planck2018}. These contributions are therefore
entirely negligible. This crucial result implies that the baryon
fluctuations generated by our model are not a modulation of the existing
adiabatic mode, but are a pristine source of new, statistically independent
fluctuations. Consequently, the fluctuations we have calculated are a pure
\textquotedblright baryon isocurvature mode\textquotedblright , where $%
\delta _{iso}(r)=\delta _{b}(r)$, uncorrelated with the primordial adiabatic
perturbations \cite{Gordon2003,Trotta2001,Lewis2000}.

In the $\Lambda $CDM framework, the total matter density contrast, $\delta
_{m}$, is the weighted sum of the cold dark matter ($\delta _{c}$) and
baryon ($\delta _{b}$) components: $\delta _{m}=f_{c}\delta _{c}+f_{b}\delta
_{b}$, where $f_{c}\approx 0.84$ and $f_{b}\approx 0.16$ \cite{Planck2018}.
Since the adiabatic mode affects all species equally ($\delta
_{c,adia}=\delta _{b,adia}=\delta _{adia}$) and our isocurvature mode
affects only baryons, the total fluctuations for each component are $\delta
_{c}=\delta _{adia}$ and $\delta _{b}=\delta _{adia}+\delta _{iso}$.
Substituting these into the expression for $\delta _{m}$ yields 
\begin{equation}
\delta _{m}=f_{c}\delta _{adia}+f_{b}(\delta _{adia}+\delta _{iso})=\delta
_{adia}+f_{b}\delta _{iso}.  \label{eq:total_contrast}
\end{equation}%
Assuming statistical independence between the adiabatic modes and our
baryogenesis mechanism, their cross-correlation vanishes \cite%
{Gordon2003,Trotta2001,Lewis2000}. The total observable matter power
spectrum is therefore a simple sum 
\begin{equation}
P_{m}(k)=\langle |\delta _{m}(k)|^{2}\rangle
=P_{adia}(k)+f_{b}^{2}P_{iso}(k).  \label{eq:total_spectrum}
\end{equation}%
Here, $P_{adia}(k)$ is the standard, observationally-verified power spectrum
of the $\Lambda $CDM model, and $P_{iso}(k)$ is the power spectrum of the
pure baryon isocurvature mode, which is precisely the quantity $P_{\delta
}(k)$ that our simulations compute: $P_{\delta }(k)=P_{iso}(k)$. The factor
of $f_{b}^{2}\approx 0.025$ naturally suppresses the contribution of our
mechanism to the total matter power spectrum, as it only perturbs the
minority baryonic component. This final expression provides the direct
theoretical link between the output of our simulations and the cosmological
observables used to constrain isocurvature modes. Our predicted values of $%
P_{\delta ,0}$ (see Table~III) lie many orders of magnitude below the upper
bound $P_{\delta ,0}\;\lesssim \;7.2\times 10^{43}\,$R$_{H}^{3}$ resulting
from the CMB constraints (see Appendix A). This isocurvature contribution is
therefore not an observable prediction of the scenario, yet entirely
compliant with current CMB constraints. It merely demonstrates that the
model is internally consistent and compatible with cosmological constraints.
Consequently, the only phenomenologically significant prediction of the
model remains the primordial magnetic field amplitude required for
baryogenesis, which lies in the range probed by Planck analyses of
primordial magnetic fields.

\subsection{On the Gaussian Assumption}

\label{sec:gaussian}

Throughout this work the magnetic vector potential components $A_{\pm,i}$
are modelled as Gaussian random fields (see Sec.~\ref{IIIC}), so that $A_{t}=|%
\mathbf{A}_{+}-\mathbf{A}_{-}|$ follows the Maxwellian distribution~(\ref%
{eq:maxwell_At}). This is a working hypothesis rather than a derived
property, shared by the large majority of the primordial-magnetic-field
literature, both for its analytical tractability and because the leading
statistical description of causal stochastic fields is Gaussian by
construction \cite{Grasso2001,Subramanian2016,Durrer2013}; non-Gaussianity
typically enters only at second order, e.g. through the quadratic dependence
of the energy--momentum tensor \cite{Planck2016}. Since $Y_{B}(A_{t})$ is
strongly non-linear, it is nonetheless legitimate to ask whether departures
from Gaussianity -- the heavy tails of intermittent MHD turbulence, or
helical fields -- could affect the predicted asymmetry.

The key observation is that the sensitivity to the distribution tails is
strongly asymmetric, and that this asymmetry works in favour of the
robustness of the result. The Maxwellian~(\ref{eq:maxwell_At}) selected by
the constraint $\langle Y_{B}\rangle =Y_{B,\mathrm{obs}}$ is peaked towards
the lower edge $A_{t,\mathrm{low}}$ of the relevant window, on the rising
flank of $Y_{B}(A_{t})$ (see Fig.~\ref{f1}) as introduced in Sec.~\ref{VA}.
Its two tails are far from equivalent. The low tail, $A_{t}<A_{t,\mathrm{low}%
}$, falls in the region where $Y_{B}\rightarrow 10^{-19}$ (the value without
baryogenesis) and $Y_{\bar{B}}$ is negligible: redistributing probability
mass there changes $\langle Y_{B}\rangle $ only marginally and harmlessly.
The high tail, $A_{t}>A_{t,\mathrm{high}}$, is the only dangerous one,
because beyond $A_{t}\sim 10^{9}$~T\thinspace m the baryon density collapses
while the antibaryon density rises steeply (Fig.~\ref{f1}). A heavy high
tail would not merely rescale $\langle Y_{B}\rangle $: it would inject
antimatter and violate the matter-domination condition $Y_{\bar{B}%
}/Y_{B}<10^{-6}$ \cite{Canetti2012}.

Crucially, this dangerous region is precisely the one that the mechanism
already excludes. As discussed in Sec.~\ref{VA}, of the two field amplitudes
reproducing the observed baryon density (Table~\ref{tab:a0_values}), only
the lower one, $A_{t,\mathrm{low}}$, is retained, the higher one being
discarded exactly because it overproduces antibaryons. The exposure to the
high tail is therefore doubly suppressed: by the decay of $f(A_{t})$ away
from its peak, and by the fact that the physical solution sits well below $%
A_{t,\mathrm{high}}$. A non-Gaussian enhancement of the tails could thus
only strengthen antimatter production in a regime that the matter-domination
criterion has already removed, leaving the retained solution at $A_{t,%
\mathrm{low}}$ essentially unaffected. Combined with the fact that the
predicted field strength is quoted at the order-of-magnitude level ($%
B_{0}\sim 10^{10}$~T, entering only through $A_{0}\propto B_{0}$), we
conclude that a moderate reshaping of the distribution would shift the
inferred amplitude by at most a factor of a few, without affecting the
central conclusion.

Helicity is a partly separate issue. A maximally helical field corresponds
to a Batchelor index $n=2$ and modifies the inverse cascade and the
injection scale $k_{\ast}$, but does not by itself break the one-point
Gaussianity of the field. Since our analysis already brackets $n\in\{0,2,4\}$
(Tables~\ref{tab:b0_values} and~\ref{tab:p0_values}), the impact of a
helical spectral slope on the required amplitude is in effect already
spanned by our results. A fully non-Gaussian treatment -- sampling $A_{t}$
from an intermittent, heavy-tailed distribution calibrated on QCD-turbulence
simulations -- would be required to quantify the residual sensitivity
exactly. Such a study lies beyond the scope of the present work, whose aim
is to establish the predictive link within the standard Gaussian framework,
and is left for future investigation.

\section{Conclusion}

In this work, we have investigated the consequences of stochastic primordial
magnetic fields for baryogenesis in a two-brane Universe. Our analysis leads
to two main results. First, our analysis suggests that this baryogenesis
mechanism is both viable and predictive. To generate the observed baryon
asymmetry, the model requires the Universe to have been filled with
primordial magnetic fields of an amplitude $\sim 10^{10}$~T at the QCD
epoch. This value is not the result of parameter tuning; rather, it is an
essential condition derived from the model. The fact that this prediction
aligns with independent predictions from magnetogenesis models \cite%
{Brandenburg1996,Forbes2000,Boyanovsky2005,Boeckel2012,Tevzadze2012}
constitutes the strongest argument of this work. It establishes a predictive
and non-trivial bridge between exotic brane physics and standard QCD plasma
dynamics. Second, spatial fluctuations of the magnetic vector potential
generate a baryon-density power spectrum that becomes universally white
noise on large scales due to the non-linear dependence of the interbrane
transition rate on the magnetic potential. This produces a baryon
isocurvature component that is statistically independent of the primordial
adiabatic perturbations and fully compatible with current CMB bounds \cite%
{Planck2016}. Although its amplitude is too small to be observable, the
result highlights a structural feature of the scenario and its internal
consistency. Overall, our findings strengthen the theoretical connection
between primordial magnetism and baryon-number generation. Since primordial
magnetic fields are constrained -- and will be probed with increasing
precision -- through their imprints on the cosmic microwave background, the
predicted value $\sim 10^{10}$~T provides a concrete observational target
for testing this baryogenesis scenario.

\section*{Acknowledgment}

The author gratefully acknowledges Paul Morel for his support with the
computational resources and for his assistance in optimizing the numerical
codes. The author also thanks Patrick Peter for helpful discussions and
comments on an earlier version of the manuscript.

The author used AI-based language models exclusively for language editing,
bibliographic assistance, and general guidance through iterative dialogue on
presentation and numerical implementation aspects. No AI tools were used for
data analysis, result generation, or scientific interpretation. All
scientific ideas, analyses, and conclusions are solely those of the author.

\section*{Data and Code Availability}

The numerical codes used in this work are publicly available on Zenodo at
DOI: 10.5281/zenodo.18138069. The repository contains the two core codes
implementing the main numerical methods described in this article, which
constitute the technically non-trivial part of the analysis.

Auxiliary scripts used for post-processing, averaging procedures, and figure
generation rely exclusively on standard numerical operations and directly
follow the prescriptions detailed in the text. Their inclusion is therefore
not necessary for assessing the validity or reproducibility of the results.

\begin{appendix}

\section{The Isocurvature Parameter and Planck Constraints from CMB Data}

\label{appendixA}

The theoretical result of Eq. (\ref{eq:total_spectrum}) establishes a direct
connection between the baryon isocurvature fluctuations generated by our
mechanism and the observable matter power spectrum. In particular, the
induced spectrum predicted by our model, $P_{\text{iso}}(k)=P_{\delta }(k)$,
adds a new contribution to the standard $\Lambda $CDM framework. In CMB
analysis, the amplitude of isocurvature fluctuations is quantified through
the parameter \cite{Planck2018} 
\begin{equation}
\beta _{\text{iso}}(k_{0})=\frac{P_{\text{iso}}(k_{0})}{P_{\text{adia}%
}(k_{0})}\,,  \label{iso}
\end{equation}%
at the comoving pivot scale $k_{0}=0.05$ Mpc$^{-1}$ \cite{Planck2018}. This
convention allows for a direct comparison between primordial spectra and the
observable angular power spectra after transfer functions are applied.
Although our isocurvature spectrum is generated at an early epoch ($T\approx
20$ MeV), its definition in comoving coordinates ensures that the spectral
shape is preserved under linear evolution. It is therefore legitimate to
compare our prediction at $k_{0}$ with CMB constraints, provided that no
late-time non-linear magnetohydrodynamical processes significantly alter the
spectrum, an assumption justified in Sec. \ref{IIIA}. Planck 2018 data
constrain pure, uncorrelated baryon isocurvature modes to contribute less
than a few percent of the total power \cite{Planck2018}. For a nearly
scale-invariant isocurvature spectrum, the 95\% C.L. bound reads $\beta _{%
\text{iso}}\lesssim 10^{-2}$ \cite{Planck2018}. While this bound is formally
derived for scale-invariant spectra, it provides a robust order-of-magnitude
limit for our case as well, since our predicted spectrum is effectively
white noise ($P_{iso}(k)=P_{\delta }(k)\propto k^{0}$) up to the cutoff
scale $k_{\ast }$.

The adiabatic spectrum at the pivot scale is well known from the Planck
results. It is usually parametrized in terms of the dimensionless curvature
spectrum \cite{Planck2018} 
\begin{equation}
\Delta _{\text{adia}}^{2}(k)\equiv \frac{k^{3}}{2\pi ^{2}}\,P_{\text{adia}%
}(k),  \label{DR}
\end{equation}%
with amplitude at the pivot scale $k_{0}=0.05\,\text{Mpc}^{-1}$ given by 
\cite{Planck2018} 
\begin{equation}
\Delta _{\text{adia}}^{2}(k_{0})=A_{s}\simeq 2.1\times 10^{-9}.
\label{DRlim}
\end{equation}%
The baryonic isocurvature component is defined as \cite{Planck2018} 
\begin{equation}
\Delta _{\text{iso}}^{2}(k)\equiv \frac{k^{3}}{2\pi ^{2}}\,P_{\text{iso}}(k),
\label{Diso}
\end{equation}%
so that the relative isocurvature fraction at the pivot reads \cite%
{Planck2018} 
\begin{equation}
\beta _{\text{iso}}(k_{0})\equiv \frac{\Delta _{\text{iso}}^{2}(k_{0})}{%
\Delta _{\text{adia}}^{2}(k_{0})}=\frac{(k_{0}^{3}/2\pi ^{2})\,P_{\delta ,0}%
}{A_{s}}.  \label{betarel}
\end{equation}%
Since the adiabatic and isocurvature modes are generated by physically
distinct mechanisms (inflation and post-QCD baryogenesis, respectively),
their cross-correlation can be safely neglected. The spectrum in our model
is generated at $T\simeq 20$ MeV, well before recombination. However, the
CMB pivot scale $k_{0}=0.05$ Mpc$^{-1}$ remains super-horizon at such
temperatures, so that the relative amplitude of isocurvature to adiabatic
modes is conserved until horizon entry. As a consequence, no additional
transfer function between $T=20$ MeV and recombination is required for the
comparison performed above. It should be noted that this pivot scale lies
far outside the range plotted in Fig.~\ref{f2} (i.e., $k_{0}\ll k_{IR}$).
Nevertheless, the predicted white-noise nature of our spectrum for $%
k<k_{\ast }$ allows its constant amplitude, $P_{\delta ,0}$, to be robustly
extrapolated to these much larger scales.

The Planck constraints require $\beta_{\text{iso}}(k_{0})\lesssim \beta_{%
\text{limit}}$, with $\beta _{\text{limit}}\sim 10^{-2}$ depending on the
assumed mode \cite{Planck2018}. This translates into the following bound on
the amplitude $P_{\delta ,0}$ 
\begin{equation}
P_{\delta ,0}\;\lesssim \;\frac{2\pi ^{2}}{k_{0}^{3}}\,\beta _{\text{limit}%
}\,A_{s}.  \label{limit}
\end{equation}%
Numerically, for $k_{0}=0.05$ Mpc$^{-1}$, this gives at $T\simeq 20$ MeV
from the CMB data 
\begin{equation}
P_{\delta ,0}\;\lesssim \;3.3\times 10^{-6}\,\text{Mpc}^{3}\;\;\simeq
\;7.2\times 10^{43}\,\text{R}_{H}^{3},  \label{Constraint}
\end{equation}

with R$_{H} = 1.1 \times 10^{6}$ m \cite{Cline2019}.

\end{appendix}

% Bibliography

\end{document}